\documentclass[reprint,showpacs,amsfonts,amssymb,amsmath,aps,pra,superscriptaddress]{revtex4-1}

\usepackage[ascii]{inputenc}
\usepackage[T1]{fontenc}
\usepackage[english]{babel}
\usepackage{graphicx}
\usepackage{microtype}
\usepackage{bm}
\usepackage[pdftex,colorlinks]{hyperref}
\hypersetup{%
  linkcolor=blue,%
  citecolor=blue,%
  urlcolor=blue,%
  pdftitle={Purity oscillations in coupled Bose-Einstein condensates},%
  pdfauthor={Jonathan Stysch, Felix Roser, Dennis Dast, Holger Cartarius, G\"unter Wunner}%
}

\newcommand{\ket}[1]{|{#1}\rangle}

\begin{document}

\title{Purity oscillations in coupled Bose-Einstein condensates}

\author{Jonathan Stysch}
\email[]{jonathan.stysch@itp1.uni-stuttgart.de}
\affiliation{Institut f\"ur Theoretische Physik 1, Universit\"at Stuttgart,
  70550 Stuttgart, Germany}
\author{Felix Roser}
\affiliation{Institut f\"ur Theoretische Physik 1, Universit\"at Stuttgart,
  70550 Stuttgart, Germany}
\author{Dennis Dast}
\affiliation{Institut f\"ur Theoretische Physik 1, Universit\"at Stuttgart,
  70550 Stuttgart, Germany}
\author{Holger Cartarius}
\affiliation{Institut f\"ur Theoretische Physik 1, Universit\"at Stuttgart,
  70550 Stuttgart, Germany}
\affiliation{Physik und ihre Didaktik, 5.\ Physikalisches Institut,
  Universit\"at Stuttgart, 70550 Stuttgart, Germany}
\author{G\"unter Wunner}
\affiliation{Institut f\"ur Theoretische Physik 1, Universit\"at Stuttgart,
  70550 Stuttgart, Germany}

\date{\today}

\begin{abstract}
  Distinct oscillations of the purity of the single-particle density
  matrix for many-body open quantum systems have been shown to exist
  [Phys. Rev. A 93, 033617 (2016)]. They are found in
  $\mathcal{PT}$-symmetric Bose-Einstein condensates, in which the coherence
  of the condensate drops and is almost completely restored periodically.
  For this effect the presence of a gain and loss of particles turned out
  to be essential. We demonstrate that it can also be found in closed
  quantum systems of which subsystems experience a gain and loss of
  particles. This is shown with two different lattice setups for cold
  atoms, viz.\ a ring of six lattice sites with periodic boundary conditions
  and a linear chain of four lattice wells. In both cases pronounced purity
  oscillations are found, and it is shown that they can be made experimentally
  accessible via the average contrast in interference experiments.
  This shows that it is possible to identify this characteristic
  effect in closed quantum systems which have been proposed to realize
  $\mathcal{PT}$-symmetric quantum mechanics via separation into subsystems.
\end{abstract}

\maketitle

\section{Introduction}
\label{sec:introduction}

Today it is well established that $\mathcal{PT}$-symmetric quantum mechanics
\cite{Bender98a,Bender99a} can be used as an effective description of
open quantum systems \cite{Moiseyev2011a}. In this context complex potentials
introduce a coupling to an environment which is not defined in detail. A
negative imaginary part describes a decrease of the probability amplitude of a
quantum particle to be in the system under consideration. In the same sense a
positive imaginary potential represents an increase of this probability
amplitude. $\mathcal{PT}$ symmetry of the Hamiltonian offers the possibility of
balanced gain and loss, since a $\mathcal{PT}$-symmetric potential has spatially
separated sinks and sources of the probability amplitude, but with the same
strength. This was introduced and used in many quantum mechanical
applications \cite{Znojil1999a,Mehri-Dehnavi2010a,Jones2010a,Graefe2008a,%
  Musslimani2008a} as well as in quantum field theories \cite{Bender2012a,%
  Mannheim2013a,Schwarz2015a,Abt2015a}, where a rich variety of characteristic
effects of these systems, such as self-orthogonality, spontaneous
$\mathcal{PT}$ symmetry breaking or quasi-Hermiticity was uncovered. However,
the formalism is not restricted to quantum mechanics but also has
applications in classical systems such as electromagnetic waves
\cite{Bittner2012a,Klaiman08a,Wiersig2014a,Bittner2014a,Doppler2016a} or
electronic devices \cite{Schindler2011a}.

In optics in particular the concept of $\mathcal{PT}$ symmetry, or non-Hermitian
Hamiltonians in general, has been very fruitful. Based on the fact that
imaginary contributions to the refractive index can be understood as sources
and sinks of the electromagnetic field amplitude \cite{El-Ganainy2007a} systems
of optical wave guides with balanced gain and loss were constructed and 
successfully proved the applicability of the theoretical formalism in
experiments \cite{Rueter10a,Peng2014a,Guo09a}. This has led to remarkable
experimental verifications such as relations to topological properties of
lattice systems \cite{Weimann2016a} or non-adiabatic state flips
\cite{Doppler2016a}.

If a single particle is studied, the complex potentials act on the particle's
probability amplitude to be in the system \cite{Graefe10a}. A different
interpretation can be obtained in many-particle systems. For example, in
the Gross-Pitaevskii equation of a Bose-Einstein condensate (BEC) the gain and
loss terms in the potential modify the amplitude of the mean-field wave
function, and thus describe a coherent injection or removal of particles
\cite{Dast2013a}. In this approach the quantum system of a BEC is only discussed
on the mean-field level and its quantum many-body character remains hidden.

However, the many-body properties are clearly important for the system. On the
one hand it can be conjectured that quantum fluctuations make the appearance of
exact $\mathcal{PT}$ symmetry impossible \cite{Schomerus2010a}. On the other
hand the dynamics of a condensate subject to balanced gain and loss shows a
characteristic signature. As was shown \cite{Dast2016a,Dast2016b} the coherence
of a condensate can be affected substantially by the gain and loss effects. In
the system studied in Refs.\ \cite{Dast2016a,Dast2016b} the purity of the
single-particle density matrix, which can be used as a measure of
how close the condensate is to a mean-field state, drops periodically during
the dynamics but is also almost completely restored in each period.
This is in contradiction with the usual assumption that coherence gets
lost due to interactions but never recurs.

A very fundamental study of the corresponding mechanism for the information
flow in linear quantum systems is presented by Kawabata et al.\
\cite{Kawabata2017a}. Experimental studies of linear systems with respect
to the available information were done in Refs.\
\cite{Lei2018a,Naghiloo2019a,Bian2019a}.

The results of Refs.\ \cite{Dast2016a,Dast2016b} were obtained in a
many-particle description, where a master equation in Lindblad form
\cite{Breuer02a} was used to introduce the gain and loss of atoms. It is known
that this description shows all effects visible in the Gross-Pitaevskii
equation of the system and converges to its imaginary potentials in the
mean-field limit of the many-body system. However, in this description the
sources and drains of particles are still located in some environment, which
is not specified further. Since proposals exist how $\mathcal{PT}$-symmetric
BECs can be realized, we wish to investigate whether or not these realizations
exhibit the purity oscillations. We suggest to exploit such possible
realizations and in this work we demonstrate that purity oscillations can
indeed be found and identified beyond any doubt in these systems. Since
these proposals are subsystems embedded in a larger closed structure, we
show here that purity oscillations can be found in closed quantum systems,
which has, to the best of our knowledge, never been observed before. This
renders the experimental realization of quantum systems with purity
oscillations possible.

Our starting point is the proposal of Kreibich et al.\
\cite{Kreibich2013a,Kreibich2016a}, which consists of a multi-well trap for
the BEC. Some of the inner wells are considered as the system and the outer
wells form the environment. In the mean-field limit the influx and outflux of
particles into and from the system have exactly the same influence on the
condensate as the imaginary potentials. A primary objective of the present work
is to investigate whether such a Hermitian system, of which only a subsystem is
considered, can show a behavior similar to the open system with respect to the
time evolution of the coherence in the system, i.e.\ whether or not the
purity oscillations discussed by Dast et al. \cite{Dast2016a} can occur in this
Hermitian system.

To do so, we use two approaches. The first consists of a six-well setup with
periodic boundary conditions. It consists basically of two coupled trimers,
certainly a nontrivial system since already the isolated trimer can exhibit a
rich dynamics \cite{Franzosi2003a}. The second one is a chain of four wells,
which is very close to the original proposal of Kreibich et al.\
\cite{Kreibich2013a} and in which only the inner two wells are regarded as the
system with gain and loss. The many-body dynamics of these systems is studied
and the purity of the single particle density matrix is calculated. This will
show that, as in the open system \cite{Dast2016a}, due to the gain and loss of
atoms in the subsystem the purity can perform oscillations, in which it drops
to small values but always is nearly completely restored. As was discussed
by Dast et al.\ \cite{Dast2016a} this effect leads to a measurable quantity
in interference experiments since a high purity is necessary for a distinct
average contrast in interference patters. At low values of the purity the
average contrast vanishes. We confirm that this behavior is also present in
our setups.

To do so, we start with a six-well system with periodic boundary conditions in
Sec.\ \ref{sec:periodic}. After the preparation of the initial state
(Sec.\ \ref{sec:inital}) and the introduction of the calculations for the
dynamics (Sec.\ \ref{sec:mbd}) we investigate the particle number dynamics in
the single wells (Sec.\ \ref{sec:pnd}) and show the effect on the purity of
the single-particle density matrix (Sec.\ \ref{sec:purity}) as well as on the
contrast in interference experiments (Sec.\ \ref{sec:contrast}). This is
compared with the dynamics of a chain of four wells in Sec.\
\ref{sec:fourwelles}. Conclusions from these numerical studies are drawn
in Sec.\ \ref{sec:conclusion}.

\section{Dynamics of condensates with periodic boundary
  conditions}
\label{sec:periodic}
\begin{figure}
  \centering
  \includegraphics[width=0.8\columnwidth]{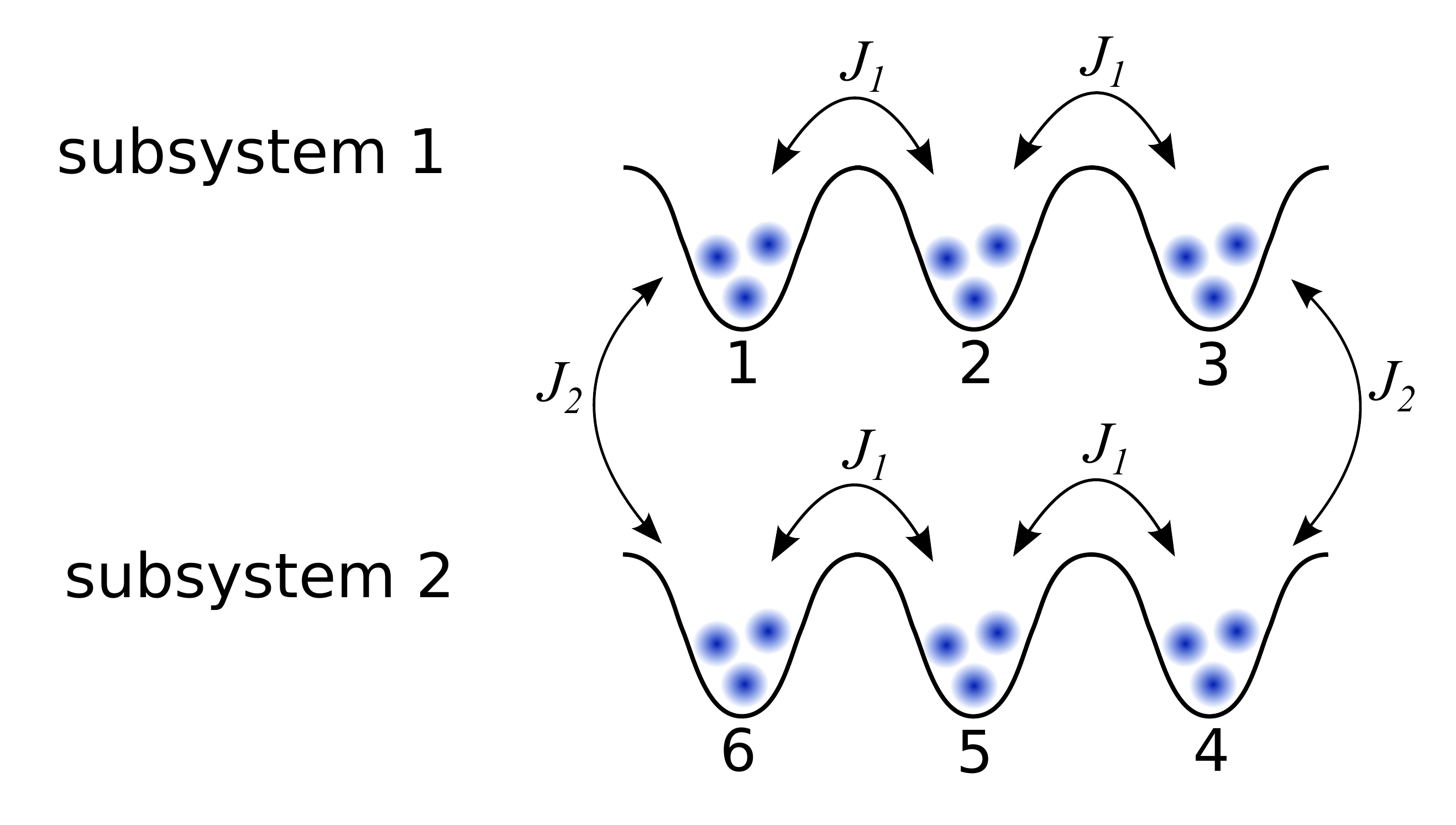}
  \caption{Schematic representation of the setup discussed in Sec.\
    \ref{sec:periodic}: Two coupled three-site subsystems forming a six-site
    Hermitian system.}
  \label{fig:system1}
\end{figure}

In this section the system proposed in Fig.~\ref{fig:system1}
is analyzed, with the central question under investigation as to what extent
the many-particle dynamics of two coupled BECs shows a behavior similar to
that of a condensate in a simple open system. A special focus is set on the
time evolution of the coherence of the condensates as expressed by the purity
of the single-particle density matrix.

\subsection{Initial state}
\label{sec:inital}

The situation considered is as follows. The two subsystems are initially fully
separated by setting the tunneling strength $J_2$ to zero. At $t=0$ each
subsystem is populated with a pure BEC. However, the two BECs do not share
phase coherence with each other. The potential wall between the subsystems is
then lowered, assigning a finite value to $J_2$ and inducing a dynamics in the
system.

An initial state representing the situation described above can be found
by proposing mean-field states for each subsystem individually, which are pure
a priori. These states can be calculated with the discrete dimensionless
Gross-Pitaevskii equation for three sites in the stationary case, i.e.\
\begin{align}
  \begin{split}
    \mu c_1  & = g|c_1|^2 c_1 -J_1c_2 \; ,\\
    \mu  c_2  & = g|c_2|^2 c_2 -J_1c_1-J_1c_3 \; ,\\
    \mu  c_3  & = g|c_3|^2 c_3 -J_1c_2 \; ,
  \end{split}
\end{align}
where $g$ represents the macroscopic particle-particle interaction and $\mu$
is the chemical potential. The ground state of this system of equations is
calculated numerically under the normalization condition $|c_1|^2+|c_2|^2
+|c_3|^2 =\frac{1}{2}$, which ensures that the combined system is of norm one.
The mean-field coefficients $c_i$ of the ground state can be chosen real at all
sites $i$ and for the exemplary case of $g=J_1=1$ the calculation yields
$c_1=c_3 \approx 0.3604$ and $c_2 \approx 0.4902$.

\subsection{Many-particle dynamics}
\label{sec:mbd}

In general the many-particle state in the Fock base corresponding to a
mean-field state is given by
\begin{multline}
  \ket{\bm{c},N}= \sum\limits_{n_1+\cdots+n_M=N}
  \sqrt{\frac{N!}{n_1!\cdots n_M!}}\\
  \times c_1^{n_1}\cdots c_M^{n_M} \ket{n_1,\dots,n_M} \; ,
  \label{eq:fockState}
\end{multline}
where $N$ is the total number of particles in the system, $M$ is the number of
sites and the coefficients $n_j$ denote the number of particles at site $j$.
For our case of two independent mean-field states in two identical subsystems
a direct product of two three-well states is needed with $N/2$ in each
subsystem, i.e.\
\begin{equation}
  \ket{\psi} = \ket{\{c_1,c_2,c_3\},N/2} \otimes \ket{\{c_4,c_5,c_6\},N/2} \; .
\end{equation}
The coefficients $c_1 = c_3 = c_4= c_6$ and $c_2 = c_5$ are those calculated
above.

The dynamics of the many-particle system is solved with a Bose-Hubbard type
Hamiltonian
\begin{equation}
  H = -\sum_{i,j} J_{ij} a^\dagger_i a_j + \frac{U}{2}\sum_{j} U a^\dagger_j
  a^\dagger_j a_j a_j 
  \label{eq:bhh}
\end{equation}
with $J_{ji} = J_{ij}$, in which no on-site energy term is present since all
wells are assumed to have the same on-site energy and dimensionless units are
chosen appropriately. The operators $a_i^\dagger$ and $a_i$ create and
annihilate a particle at site $i$, respectively. To reflect the system of Fig.\
\ref{fig:system1} and to agree with the mean-field state used as initial
condition we set $J_{12} = J_{23} = J_{45} = J_{56} = J_1$, $J_{34} = J_{61}
= J_2$, and $U = g/(N-1)$.

\subsection{Filling level dynamics}
\label{sec:pnd}

Central quantities in analyzing the dynamical behavior are the expectation
values of the number operator $\langle\hat{n}_j\rangle
= \langle\hat{a}_j^\dagger\hat{a}_j\rangle$ for each site (referred to as
filling levels in the following). Because of the system's symmetry that
of the initial state does not get lost during the temporal evolution and there
are only two independent filling levels, viz.\ the central levels of the
subsystems ($c_2=c_5$), and all other levels ($c_1 = c_3 = c_4 = c_6$).

\begin{figure}
  \includegraphics[width=\columnwidth]{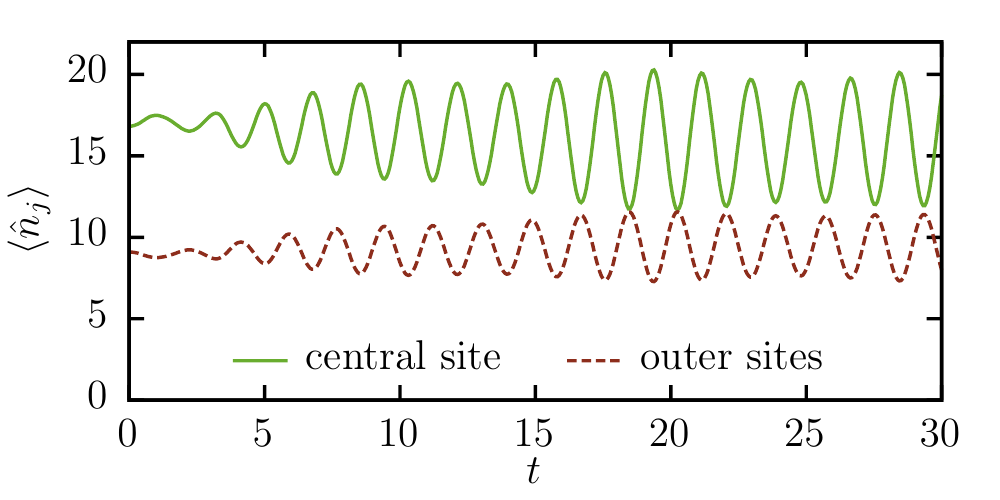}
  \caption{Time evolution of the filling levels for $N=70$, $g=J_1=1$ and
    $J_2=2$ for two initially fully separated BECs. Relatively stable
    oscillations are established after some on-set period. Due to the symmetry
    of the system and conservation of the total particle number, there is a
    phase difference of $\pi$ between the oscillations of the central site
    (solid line) and those of the outer sites (dashed line).}
  \label{fig:J2g1N}
\end{figure}

By propagating the initial state with a simple Runge-Kutta algorithm the time
evolution of the many-particle states is calculated for a system of $N= 70$
particles. Of course the system's dynamics depends critically on its
parameters. A detailed study of the parameter dependence in the description via
a master equation is given in Refs.\ \cite{Dast2016b,Dast2017a}. In this paper
we wish to observe clear purity oscillations, and thus search for appropriate
parameters. The nonlinear interaction strength $g$ and the tunneling strength
within the subsystems $J_1$ are set to one as above and the behavior of the
system is investigated for different couplings $J_2$ between the subsystems
since this parameter induces the dynamics after the condensates have been
prepared independently. In our study a value of $J_2=2$ turned out to provide a
dynamics very similar to that of the investigation of open systems
\cite{Dast2016a}. An example for this choice of $J_2$ is shown in Fig.\
\ref{fig:J2g1N}.
After some on-set period relatively stable oscillations are established. Due
to the underlying symmetry they show a phase difference of $\pi$ between the
central site of each subsystem and its outer sites. A stronger coupling of the
subsystems (i.e.\ $J_2 > 2J_1$) also yields an oscillatory behavior. However,
the amplitude and stability of the oscillations decrease. Larger choices of
the nonlinearity $g$ result in very unstable oscillations as well.

\subsection{Investigation of the purity}
\label{sec:purity}

In accordance with the study in open systems \cite{Dast2016a} we use the
purity
\begin{equation}
  P = \frac{M}{M-1} \mathrm{tr}\, \left ( \bm{\sigma}_\mathrm{red}^2 \right )
  - \frac{1}{M-1}
  \label{eq:purity}
\end{equation}
of the reduced single-particle density matrix $\bm{\sigma}_\mathrm{red}$ with
the elements
\begin{equation}
  \sigma_{\mathrm{red},ij} = \frac{\langle a_i^\dagger a_j \rangle}{\sum_{k=1}^M
    \langle a_k^\dagger a_k \rangle}
\end{equation}
and the dimension $M$ of the system to quantify how close the state considered
is to a pure condensate. One of the subsystems corresponds to the open system
of Ref. \cite{Dast2016a}, and thus we use its purity $P$ with $i,j\in \{1,2,
3\}$ and $M=3$. For completeness we compare it with the purity $P_\mathrm{tot}$
of the whole system, i.e.\ $P_\mathrm{tot}$ is calculated with $i,j\in \{1,
\dots,6\}$ and $M=6$.

\begin{figure}
  \includegraphics[width=\columnwidth]{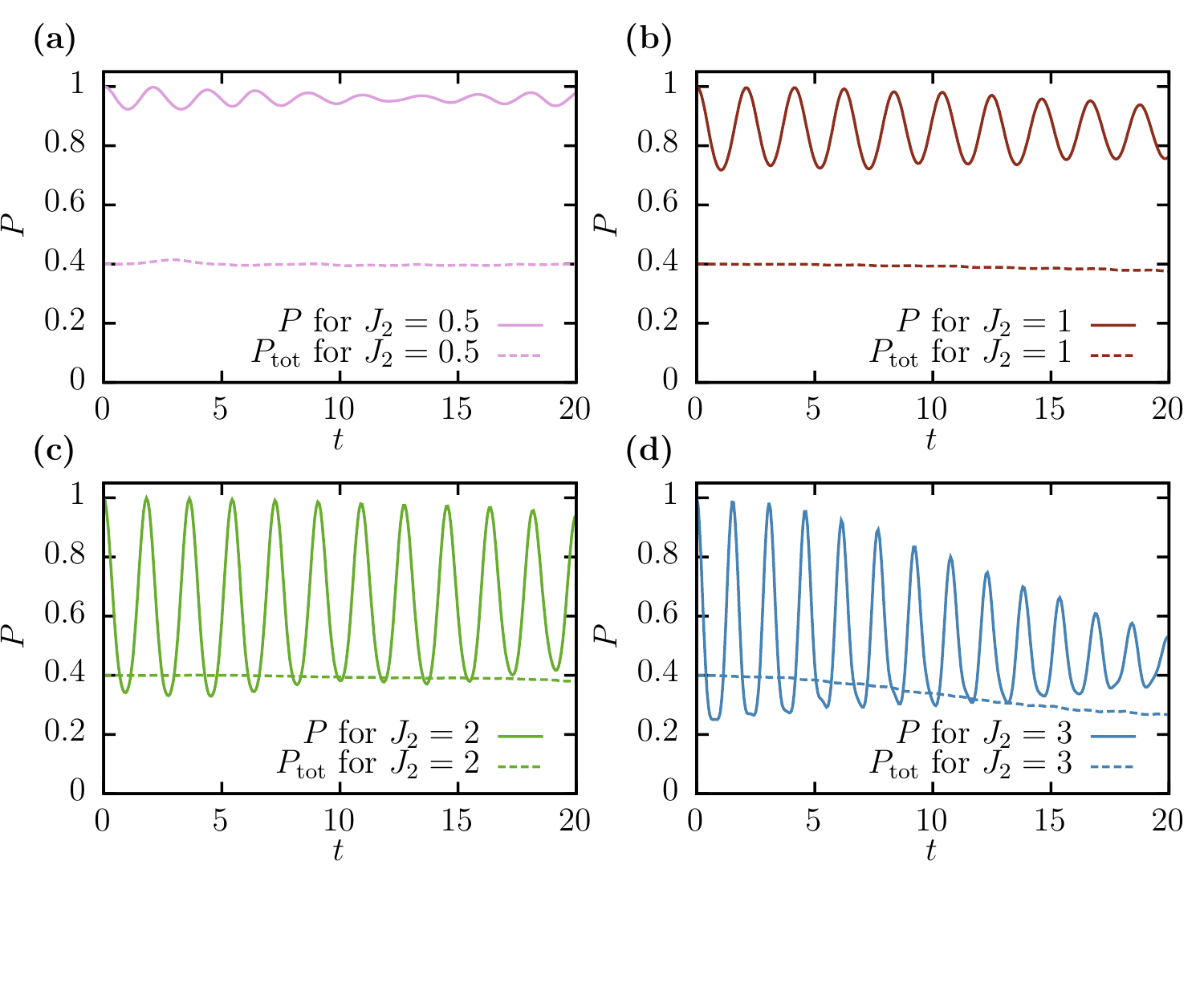}
  \caption{Time dependence of the purity $P$ (solid lines) of one subsystem
    and the purity $P_\mathrm{tot}$ (dashed lines) of the total system for
    different values (a) $J_2 = 0.5$, (b) $J_2 = 1$, (c)  $J_2 = 2$, and (d)
    $J_2 = 3$ of the coupling strength between the two systems. The other
    parameters are $N=70$ and $g=J_1=1$. The purity $P$ undergoes oscillations
    that can be very distinct and of a large amplitude for higher values of
    $J_2$. The total purity stays almost constant at $P_\mathrm{tot} = 0.4$ and
    decreases only slightly over time, which can be explained solely by
    statistical reasons. The amplitude and average of the oscillations of $P$
    decrease over time as well, especially for the highest value of the
    tunneling strength $J_2=3$ (d).}
  \label{fig:JCompP}
\end{figure}

Since it is known that the purity oscillations are strongly related with the
coupling of a system to its gain-loss environment \cite{Dast2016a,Dast2016b}
we investigate the purity for different strengths of the coupling of the two
condensates as mediated by the tunneling constant $J_2$ in Fig.\
\ref{fig:JCompP}.
The purity $P$ (solid lines) of one subsystem is displayed alongside the
overall purity $P_\mathrm{tot}$ (dashed lines) of the six-site system. Since
the subsystems are initially prepared in a mean-field state, i.e.\ with a
pure condensate, all curves in the figure have to start at $P=1$ for $t=0$.
As time progresses the purity does in fact undergo oscillations, whose
amplitude is strongly determined by the strength of the coupling.
For $J_2 = 0.5$ [Fig.\ \ref{fig:JCompP}(a)] the minimum of the first
oscillation is still higher than $P = 0.9$. It decreases for the stronger
coupling $J_2 = 1$ [Fig.\ \ref{fig:JCompP}(b)] and reaches a value of
$P\approx 0.35$ for $J_2  =2 $ [Fig.\ \ref{fig:JCompP}(c)].

Increasing the coupling even further to $J_2 = 3$ [Fig.\ \ref{fig:JCompP}(d)]
does only result in a slight initial increase of the amplitude. This minimal
improvement goes along with an unwanted strong overall decrease of the purity
with time. In addition, the almost harmonic shape that the oscillations possess
for smaller values of $J_2$ is lost. This is in agreement with the study
of the open system \cite{Dast2016a,Dast2016b}, in which it was found that the
purity oscillations are aligned with the particle number oscillations. In our
closed system the most pronounced purity oscillations with an almost constant
amplitude are achieved for the coupling strength $J_2 = 2$, which also leads
to the most stable particle number oscillations.

As mentioned above, previous studies \cite{Dast2016a,Witthaut2008a} found the
rapid loss and subsequent restoration of the coherence to be a phenomenon of
open systems, which is not possible without the coupling to an environment.
In full agreement, in our case it is only possible to observe such purity
oscillations by evaluating the purity for the two subsystems individually in
the coupled Hermitian system presented in this paper. That is, for one
subsystem the other assumes the role of the environment.

Since the initial state of the system consists of two completely separated and
incoherent BECs the overall purity starts below one at $P_\mathrm{tot} = 0.4$,
which corresponds to two states with nonzero occupation of magnitude $N/2$.
These are exactly the two separate mean-field states. As time progresses
$P_\mathrm{tot}$ does not show any dynamical behavior except for a slight
decrease due to statistical reasons (i.e. there are more accessible states of
lower purity). An overall deterioration of the subsystem's purity $P$ can also
be observed in the long-term time evolution for the same reasons. On a long
time scale the amplitude as well as the average value of the oscillations
decrease.

\subsection{Contrast in interference experiments}
\label{sec:contrast}

The coherence of the atoms in a BEC as measured by the purity plays a crucial
role in interference experiments. As demonstrated e.g.\ in Ref.\
\cite{Andrews1997a}, the potential barrier between two lattice sites can be
turned off, which results in an expansion of the atomic clouds of each site
such that they ultimately interfere. An interference pattern can be visualized
with a light source and detected using a CCD camera. For a system of low
coherence these interference patterns will be different each time the
interference experiment is executed because in this case there is no defined
phase between the atoms of each site. However, if the system is coherent and
there is a defined phase relation between the atoms of both sites, the
interference pattern will be identical if the experiment is repeated under the
same conditions. This behavior is expressed in the average contrast of the
interference pattern. For the pattern created by the atoms of two neighboring
sites $j$ and $k$ this term can be expressed in terms of the elements of the
single-particle density matrix \cite{Witthaut2008a},
\begin{equation}
  \nu_{jk} = \frac{2|\! \langle\hat{a}_j^\dagger\hat{a}_k\rangle\!|}
  {\langle\hat{a}_j^\dagger\hat{a}_j\rangle+\langle\hat{a}_k^\dagger
    \hat{a}_k\rangle} \; \; \in\, [0,1] \; .
\end{equation}

The coherence between both sites is quantified by the two-site purity
\begin{equation}
  P_{jk} = \frac{\left (\langle \hat{a}_j^\dagger \hat{a}_j \rangle
    - \langle \hat{a}_k^\dagger \hat{a}_k \rangle \right )^2
  + 4 \langle \hat{a}_j^\dagger \hat{a}_k \rangle \langle \hat{a}_k^\dagger
  \hat{a}_j \rangle}{\left (\langle \hat{a}_j^\dagger \hat{a}_j \rangle + \langle
      \hat{a}_k^\dagger \hat{a}_k \rangle \right )^2} \; ,
\end{equation}
which is gained by considering only the matrix elements of site $j$ and $k$ in
Eq.\ \eqref{eq:purity}. Together with the squared particle imbalance
\begin{equation}
  I_{jk} = \left ( \frac{\langle \hat{a}_j^\dagger \hat{a}_j \rangle - \langle
      \hat{a}_k^\dagger \hat{a}_k \rangle}{\langle \hat{a}_j^\dagger \hat{a}_j 
      \rangle + \langle \hat{a}_k^\dagger \hat{a}_k \rangle} \right )^2
\end{equation}
this two-site purity determines the average contrast,
\begin{equation}
\nu_{jk}^2 = P_{jk} - I_{jk} \; .
\end{equation}
Hence, the two-site purity is equal to the squared average contrast if the
filling levels are identical for both sites. In general, any finite particle
imbalance lowers the average contrast while the two-site purity provides an
upper limit for $\nu_{jk}^2$.

For a system of only two sites, as e.g.\ in Refs.\ \cite{Dast2016a,Dast2016b},
the abstract quantity purity of the system is thereby related to a quantity
observable in experiment. Even though it is not possible to generalize this
concept for multiple sites in an easy manner, it is still feasible for the
system presented in this section because both outer sites of the subsystems
behave identically. Thus, if a high coherence is observed between the central
site and one of the outer sites, there has to be a high coherence in the
subsystem as a whole.

\begin{figure}
  \includegraphics[width=\columnwidth]{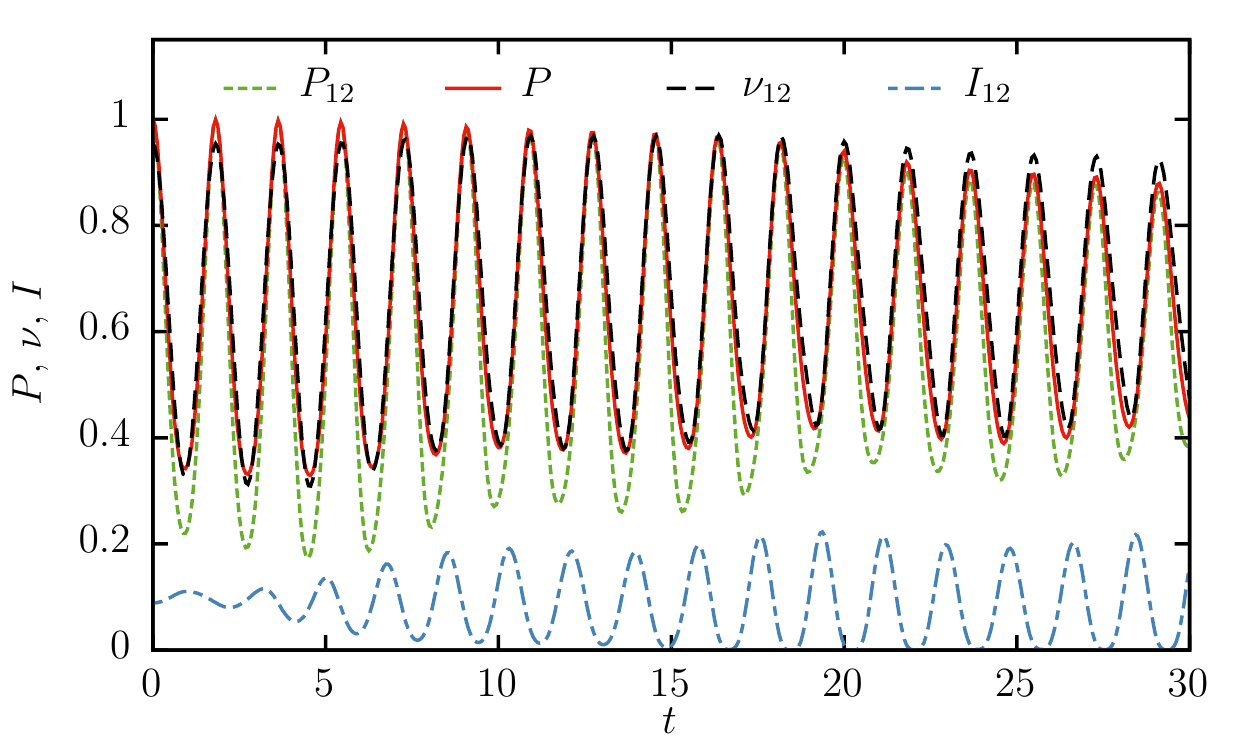}
  \caption{Average contrast $\nu_{12}$ after time $t$ in an interference
    experiment of sites 1 and 2 compared with the purity $P_{12}$, the
    squared particle imbalance $I_{12}$ of the two sites, and the overall
    purity of the subsystem $P$ for the same parameters as in Fig.\
    \ref{fig:J2g1N}. The average contrast undergoes distinct oscillations
    close in shape and size to those of the purities $P_{12}$ and $P$. The
    moderate particle imbalance is not able to disturb this behavior, partly
    since its minima almost coincide with the maximum of the purity.
  }
  \label{fig:AC}
\end{figure}

However, for the average contrast to be largely determined by the purity, the
dynamics of the particle imbalance has to play only a minor role. To explore
the relationship of these quantities for the system under investigation the
time evolution of the average contrast $\nu_{12}$ in an interference
of sites 1 and 2 is compared in Fig.~\ref{fig:AC},
with the particle imbalance $I_{12}$, the two-site purity $P_{12}$, and the
overall purity of the subsystem $P$. Right from the start of the time
evolution, while the dynamics of the filling levels are still in their
on-set phase, there are already distinct oscillations of the average contrast
closely related in frequency and shape to the oscillations of the purity.

Although $\nu_{12}$ is smaller than one for $t=0$ this small discrepancy caused
by the finite particle imbalance does not effect the qualitative behavior.
Furthermore, the maxima and minima of the evolving oscillations of $I_{12}$
almost coincide with the minima and maxima of the purity, respectively.
Therefore the dynamical behavior of the particle imbalance does not compromise
the alignment of the oscillations of the average contrast and the purity, but
rather reinforces it. This beneficial phase relation between the purity and
the particle imbalance is due to the strict symmetries of the system and cannot
be expected to occur generally.

\section{Dynamics of condensates in a chain of four wells}
\label{sec:fourwelles}

\begin{figure}
  \centering
  \includegraphics[width=0.8\columnwidth]{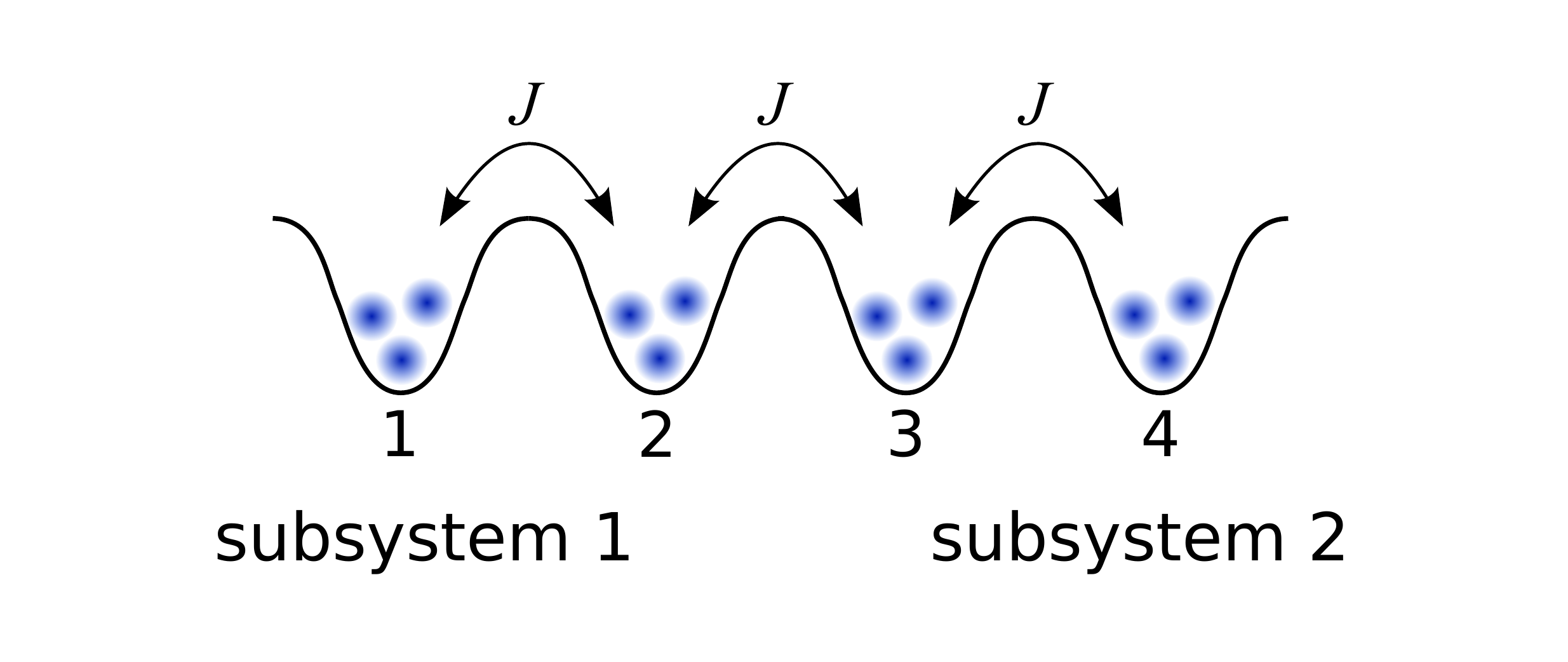}
  \caption{Schematic representation of the chain setup, in which four lattice
    sites are coupled. The left pair of wells forms subsystem 1 and the right
    pair establishes subsystem 2.}
  \label{fig:system_chain}
\end{figure}

Instead of the system shown in Fig.\ \ref{fig:system1}, one can also consider
a chain of wells without periodic boundary conditions. These systems can be
set up in a line, and thus can be realized in a simpler way as compared
to the ring studied in the last section. However, in this case the periodic
boundary conditions are lost. As we will see in the following, this is not
necessarily a drawback for the observation of purity oscillations in general.
We were able to identify them in a chain of six wells, built similarly to the
setup shown in Fig.\ \ref{fig:system1} by simply opening the chain. However,
the loss of symmetry in the system leads to a more irregular dynamics.
Intuitively, smaller systems with fewer degrees of freedom should show a less
complicated dynamics. To put this assumption to a test, we study a chain of
only four lattice sites with one condensate in the left pair of wells
(subsystem 1) and another condensate in the right pair of wells (subsystem 2)
as shown in Fig.\ \ref{fig:system_chain}.
To calculate the dynamics of this system, Eq.\ \eqref{eq:bhh} can be adapted
by taking the sum over only four sites and loosing the coupling between the
first and last site, which would close the ring.

\begin{figure}
  \includegraphics[width=\columnwidth]{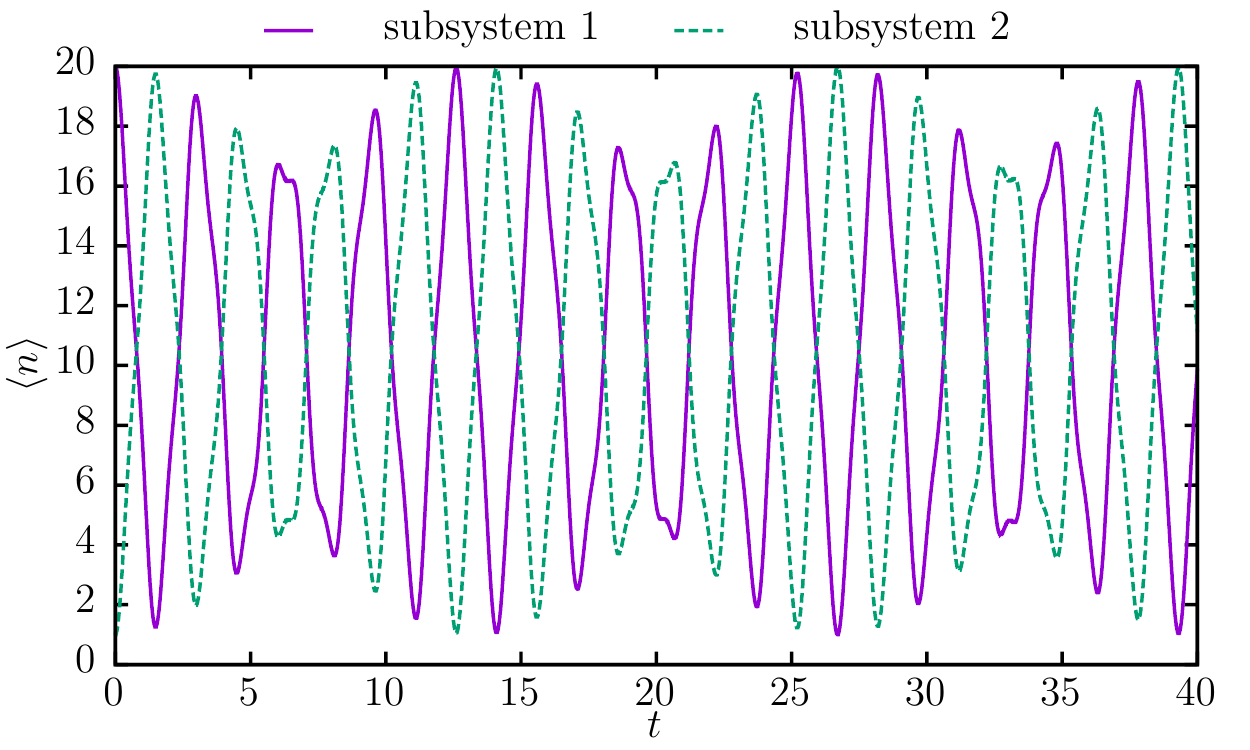}
  \caption{Particle numbers in the subsystems of the chain of four wells. The
    initial state consists of 20 particles in subsystem 1 and only one
    particle in subsystem 2. For a coupling constant $J=2$ and no
    particle-particle interactions ($g=0$), the dynamics of the system shows
    very uniform oscillations.}
  \label{fig:chain_num}
\end{figure}

As in the setup used before, the most pronounced purity oscillations are
achieved for a coupling constant $J=2$ between the neighboring lattice sites.
It can be intuitively understood that strong oscillations of the occupation
numbers of the sites occur for high particle imbalances in the initial state.
The oscillations for these parameters are shown in Fig.\ \ref{fig:chain_num}
for the case of $g=0$, which leads to very long-lived oscillations.
As can be seen they are very pronounced and still smooth enough that one can
expect visible purity oscillations. Note that in the linear case the choice
of the interaction strength $J$ just fixes the scale of time for given atomic
parameters.

\begin{figure}
  \includegraphics[width=\columnwidth]{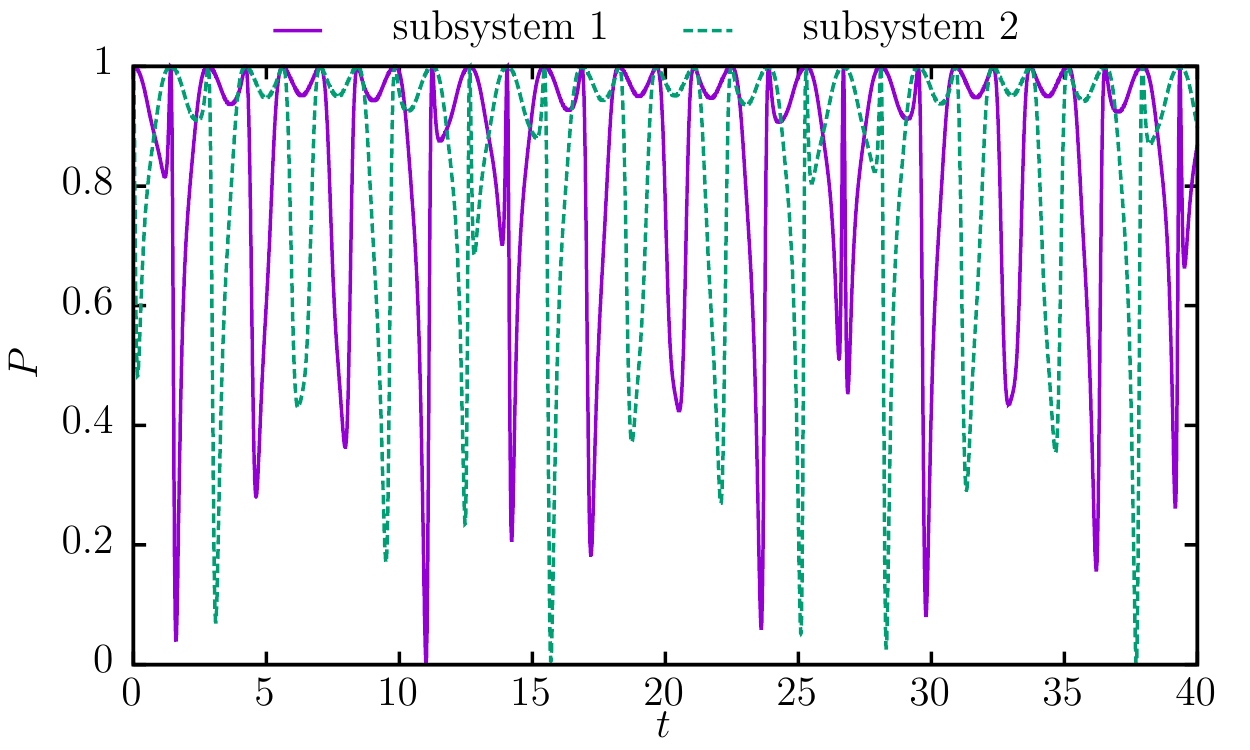}
  \caption{Purity oscillations in the two subsystems of the chain of four
    wells for the same parameters as in Fig.\ \ref{fig:chain_num}. The
    oscillations in both halfs of the system alternate with a phase shift of
    $\pi$.}
  \label{fig:chain_pur}
\end{figure}

Indeed we find that, similarly to the ring of wells discussed before, the
oscillations of the particle numbers correlate strongly with the observed
purity oscillations. In Fig.\ \ref{fig:chain_pur}
one can see that the oscillations of the purity show large amplitudes as well.
In particular, the effect that the purity is restored almost completely after
it dropped survives in the system without periodic boundary conditions.
Because of the large initial particle imbalance and the fact that the particles
do not interact with each other ($g=0$), it is obvious that the total purity
of the system remains constant at a relatively high value, which could be
confirmed in our numerical calculations.

\begin{figure}
  \includegraphics[width=\columnwidth]{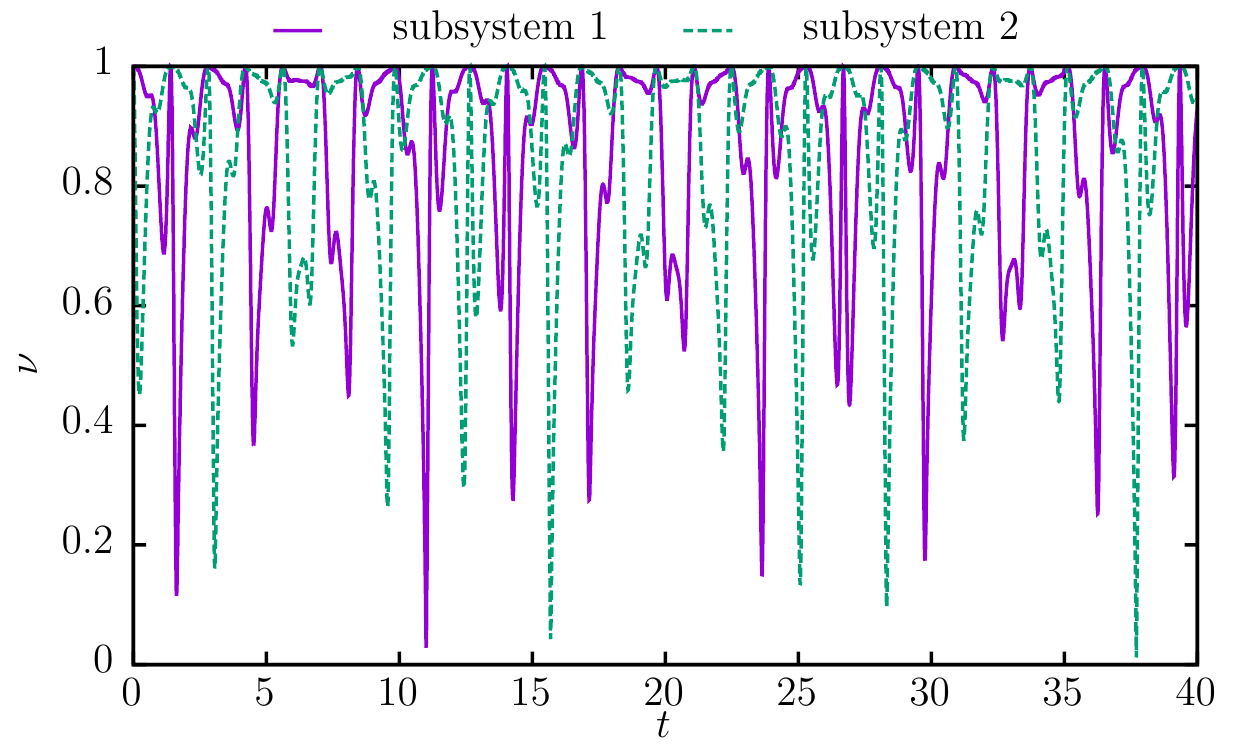}
  \caption{Average contrast in the two subsystems of the chain of four wells
    for the same parameters as in Fig.\ \ref{fig:chain_num}. The contrast
    shows a dynamic behavior similar to the purity oscillations shown in
    Fig.\ \ref{fig:chain_pur}.}
  \label{fig:chain_con}
\end{figure}

As has been shown in Sec.\ \ref{sec:contrast}, the purity oscillations of a
system influence the average contrast $\nu$, and thus the contrast can be
used to verify their existence in an experiment with atoms. In the case of our
chain the contrast also shows strong oscillations, which is confirmed in
Fig.\ \ref{fig:chain_con}.
However, we observe an increasingly irregular behavior, which can be traced back
to the more irregular behavior of the purity oscillations. Nevertheless the
purity oscillations in a chain of four wells can still be observed by
measuring the average contrast in one of the subsystems. This shows that in
principle the periodic boundary conditions are not necessary, and the
qualitative effect, viz.\ the appearance of the purity oscillations, is
equivalent in both systems. The decisive reason is the gain and loss of
particles in one of the subsystems. This can be achieved in a sufficiently
controlled manner in both setups.

\section{Conclusion}
\label{sec:conclusion}

We have demonstrated the appearance of purity oscillations in two possible
setups of multi-well systems for cold atoms, of which subsystems can be
considered as open many-body arrangements allowing for particle exchange. One
is a two-dimensional layout of six wells with periodic boundary conditions.
The second is a linear chain of four wells. In both cases clear and pronounced
purity oscillations are found. In contrast to the previous report
\cite{Dast2016a} the gain and loss of particles is not shifted to an undefined
environment but gained from a closed system and its division into subsystems.

In interference experiments the abstract purity of the single-particle density
matrix becomes a quantity indirectly observable due to its clear relation to
the average contrast. Because of the symmetric behavior of the outer sites in
the setups used in this paper the purity becomes accessible in an interference
of two neighboring sites of a subsystem. The symmetry is even such that the
particle imbalance supports the oscillations of the average contrast rather
than impairing it. Thus, we are convinced that one of these setups could
turn out to be an experimentally feasible way of demonstrating the distinct
purity oscillations predicted for balanced open quantum systems beyond the
mean-field description.

\end{document}